\documentclass{spie}
\usepackage{amsmath}
\usepackage{amsfonts}
\usepackage{amssymb}
\usepackage{mathtools}
\usepackage{commath}
\usepackage{animate}
\usepackage{microtype}
\usepackage{booktabs}
\usepackage{iftex}
\usepackage{enumitem}
\usepackage{graphicx}
\usepackage[colorlinks=true, allcolors=blue]{hyperref}
\graphicspath{{imgs/}}

\pdfpageattr {/Group << /S /Transparency /I true /CS /DeviceRGB>>}

\title{Finding novelty with uncertainty}

\author[a]{Jacob C. Reinhold}
\author[a]{Yufan He}
\author[b]{Shizhong Han}
\author[b]{Yunqiang Chen}
\author[b]{Dashan Gao}
\author[c]{Junghoon Lee}
\author[a,d]{Jerry L. Prince}
\author[a,d]{Aaron Carass}
\affil[a]{Department of Electrical and Computer Engineering, Johns Hopkins
University, Baltimore,~MD,~USA~21218}
\affil[b]{12 Sigma Technologies, San Diego,~CA~USA~92122}
\affil[c]{Department of Radiation Oncology, Johns Hopkins School of Medicine, 
Baltimore,~MD,~USA~21287}
\affil[d]{Department of Computer Science, Johns Hopkins University,
Baltimore,~MD,~USA~21218}

\authorinfo{Further author information: (Send correspondence to J.C.R.)\\J.C.R.: E-mail: jacob.reinhold@jhu.edu}

\begin{document} 
\maketitle
\begin{abstract}
Medical images are often used to detect and characterize pathology and disease;
however, automatically identifying and segmenting pathology in medical images is 
challenging because the appearance of pathology across diseases varies widely. 
To address this challenge, we propose a Bayesian deep learning method that learns to 
translate healthy computed tomography images to magnetic resonance images and 
simultaneously calculates voxel-wise uncertainty. Since high uncertainty occurs in 
pathological regions of the image, this uncertainty can be used for unsupervised 
anomaly segmentation. We show encouraging experimental results on an unsupervised anomaly 
segmentation task by combining two types of uncertainty into a novel quantity we 
call \emph{scibilic} uncertainty.
\end{abstract}
\keywords{Unsupervised anomaly segmentation, image translation, uncertainty quantification}

\section{Introduction}
\label{sec:intro}
When pathology is present in structural medical images, such as computed 
tomography~(CT) and magnetic resonance~(MR), machine learning methods can 
be used to segment the specific pathology~\cite{ronneberger2015unet}; however, 
reliable machine learning-based segmentation currently requires a labeled dataset 
of that pathology. But labeled datasets do not exist for all diseases, so having 
a machine learning method that does not rely on pathology-specific labels is desirable. 
In this paper, we propose a novel, simple-to-implement method of unsupervised 
anomaly segmentation that uses estimates of \emph{epistemic} and 
\emph{aleatoric} uncertainty~\cite{der2009aleatory} to find anomalies (i.e., novelty) 
in medical images. Our method is competitive with state-of-the-art methods.

The goal of unsupervised anomaly segmentation is to use unlabeled data (i.e., data
without ground-truth segmentation labels) to train a model that automatically
segments anomalous regions of a new image. Generally, prior methods have relied on 
training either an autoencoder~\cite{zimmerer2018context,uzunova2019unsupervised} 
or a generative adversarial network~\cite{goodfellow2014generative,schlegl2017unsupervised,baur2018deep} 
on only healthy data and using the voxel-wise reconstruction error as a heatmap for anomaly 
segmentation~\cite{an2015variational}. The underlying principle is: since the
network has only ever seen healthy data, reconstruction will be poor in regions of 
anomaly. However, unsupervised anomaly segmentation relying on reconstruction error 
perform poorly, motivating the need for alternative approaches.

Uncertainty estimation in deep neural networks~(DNN) for image translation, segmentation, and super-resolution
has been explored~\cite{bragman2018uncertainty,nair2018exploring,tanno2019uncertainty};
however, in this work, we use the uncertainty estimates to do unsupervised anomaly 
segmentation. While a limited version of this has been explored by Pawlowski et 
al.~\cite{pawlowski2018unsupervised}, we introduce a novel combination of uncertainty 
measures to increase the detection rate of anomalies as well as improve anomaly segmentation.

To estimate uncertainty, we used the work of Gal and Ghahramani~\cite{gal2016dropout}
who showed that dropout~\cite{srivastava2014dropout} can be used to learn an
approximate distribution over the weights of a DNN\textemdash a form of Bayesian inference. 
Then, during predicition, dropout is used in a Monte Carlo 
framework to draw weights from this fitted approximate distribution. The sample variance 
of the output from several stochastic forward passes corresponds to uncertainty 
about what the model knows (\emph{epistemic} uncertainty). We also modified the 
network architecture to create an additional output that corresponds to a 
variance parameter, which is fit by changing the loss function~\cite{kendall2017uncertainties}; 
this corresponds to the uncertainty about the measurement quality (\emph{aleatoric} uncertainty). 

We modified a state-of-the-art supervised image translation 
DNN\textemdash a U-Net \cite{ronneberger2015unet}\textemdash 
to capture both types of uncertainty in a CT-to-MR image translation task. The
CT-to-MR task is explored here because CT is far more common than MR images, 
but soft tissue anomalies are harder to detect in CT images. The underlying
principle proposed in this paper, however, should apply broadly to reconstruction or 
synthesis tasks; it is not specific to the CT-to-MR synthesis task. We show 
that our modified U-Net produces uncertainty estimates that can be 
used to competitively detect and segment anomalies through a combined quantity we
call \emph{scibilic} uncertainty.

\section{Methods}
\label{sec:methods}
In this section, we describe 1)~the relevant uncertainty estimation
theory and 2)~our modifications to a U-Net to estimate uncertainty.

\subsection{Uncertainty estimation}
\label{ssec:uncertainty}

Following the work of Gal~\cite{gal2016uncertainty}, we used the variance of the 
predicition as a proxy for predictive uncertainty. Predictive uncertainty can 
be split into two discrete, interpretable forms which separately estimate 
epistemic and aleatoric uncertainty. In the following paragraphs we will 
mathematically define the predictive uncertainty from which epistemic and
aleatoric uncertainty are derived.

Let our training data be denoted as
$ \mathcal D = \left\{(\mathbf x_i,\mathbf y_i) \mid \mathbf x_i, \mathbf y_i \in \mathbb{R}^M, i \in \{1,2,\ldots,N\} \right\}, $
where each $(\mathbf x_i, \mathbf y_i)$ pair are images\textemdash
co-registered MR and CT images, in our case\textemdash of flattened
length $M$. Our goal was to fit the predictive distribution,
\begin{equation}
\label{eq:preddist}
 p(\mathbf y^*\,|\, \mathbf x^*, \mathcal D) = \int p(\mathbf y^*\,|\,\mathbf x^*, \mathbf W)p(\mathbf W\,|\,\mathcal D)\,\dif \mathbf W,
\end{equation}
where $(\mathbf x^*, \mathbf y^*) \not\in \mathcal D$ is test data and $\mathbf W$
are the weights of a (multi-task) neural network $f^{\mathbf W}(\cdot)$. The
fitted predictive distribution gave us estimates of the predictive mean and 
variance, which are the synthesized image and predictive uncertainty, respectively. 

To obtain the predictive distribution, however, we needed to estimate the posterior 
$p(\mathbf W\,|\,\mathcal D)$. This posterior probability distribution over 
the weights allows us to determine epistemic uncertainty, since we expect the model 
to produce consistent outputs for known regions (i.e., low sample variance) and 
inconsistent outputs for unknown regions (i.e., high sample variance). 
But estimating the posterior distribution is not feasible due to the intractable 
integral implicit in the denominator after applying Bayes rule. 

To avoid this problem, we used variational inference~\cite{jordan1999introduction} to fit an 
approximate distribution, $q_\theta(\mathbf W)$, to $p(\mathbf W\,|\, \mathcal D)$ via optimization.
The approximate distribution was fit by minimizing the Kullback-Leibler (KL) divergence between 
$q_\theta(\mathbf W)$ and $p(\mathbf W\,|\,\mathcal D)$ using the work of Gal and Ghahramani~\cite{gal2016dropout}. 
They showed that using dropout on the full set of weights of the DNN, $\theta$, during 
normal DNN training minimizes $\mathrm{KL}(q_\theta(\mathbf W)\,||\,p(\mathbf W\,|\, \mathcal D))$. 
We used this method of variational inference to get $q_\theta^*(\mathbf W)$, 
the \emph{fitted} approximate distribution. 
We then sampled the model weights, $\widehat{\mathbf W} \sim q_\theta^*(\mathbf W)$,
to estimate the predictive distribtution, i.e., $q_\theta^*(\mathbf W)$ replaced $p(\mathbf W\,|\, \mathcal D)$ in Eq. (\ref{eq:preddist}).

To find a voxel-wise estimate of aleatoric uncertainty, we assumed the likelihood has the following form,
\begin{equation}
\label{eq:likelihood}
p(\mathbf y\,|\,\mathbf x,\mathbf W) = \mathcal N(\mathbf y; \mathbf{\hat{y}}, \mathrm{diag}(\boldsymbol{\hat{\sigma}})^2),
\end{equation}
where $\hat{\mathbf y} = f^{\mathbf W}_{\mathbf{\hat{y}}}(\mathbf x)$
and $\boldsymbol{\hat{\sigma}}^2 = f^{\mathbf W}_{\boldsymbol{\hat{\sigma}}^2}(\mathbf x)$ 
are each outputs of our multi-task neural network. Taking the negative logarithm of Eq. (\ref{eq:likelihood}) 
results in a modified form of the mean square error for the loss function,
\begin{equation}
\label{eq:loss}
\mathcal{L}(\mathbf{y}, \hat{\mathbf{y}}) = \frac{1}{M} \sum_{i=1}^M
\frac{1}{2} \boldsymbol{\hat{\sigma}}_i^{-2} \norm{\mathbf{y}_i -
\mathbf{\hat{y}}_i}_2^2 + \frac{1}{2} \log \boldsymbol{\hat{\sigma}}_i^2.
\end{equation}
When we learn the weights of the DNN according to Eq. (\ref{eq:loss}),
we are doing maximum likelihood estimation not only for $\hat{\mathbf{y}}$, 
but for the parameter $\hat{\boldsymbol{\sigma}}^2$, which is a voxel-wise
estimate of the data variance\textemdash a quantity related to aleatoric uncertainty.

As a result of these assumptions and problem setup, we can approximate the predicitive 
variance of a test sample as follows:
\begin{equation*}
\mathrm{Var}(\mathbf y^*) \approx \underbrace{\frac{1}{T}
\sum_{t=1}^T \mathrm{diag}(f^{\widehat{\mathbf
W}_t}_{\boldsymbol{\hat{\sigma}}^2}(\mathbf x^*))}_{\text{aleatoric}}
 + \underbrace{\frac{1}{T} \sum_{t=1}^T f^{\widehat{\mathbf W}_t}_{\mathbf{\hat{y}}}(\mathbf x^*) 
f^{\widehat{\mathbf W}_t}_{\mathbf{\hat{y}}}(\mathbf x^*)^\top -
\left(\frac{1}{T}\sum_{t=1}^T f^{\widehat{\mathbf W}_t}_{\mathbf{\hat{
y}}}(\mathbf x^*)\right)\left(\frac{1}{T} \sum_{t=1}^T
f^{\widehat{\mathbf W}_t}_{\mathbf{\hat{y}}}(\mathbf
x^*)\right)^\top}_{\text{epistemic}}\!,
\end{equation*}
where $T$ is the number of sampled weights. 
Consequently, the epistemic uncertainty is the term in the predictive
variance that corresponds to sample variance while the aleatoric
uncertainty is the term associated with the estimated variance of the data.

To create a heatmap used to detect anomalies, we then divided the 
epistemic uncertainty by the aleatoric uncertainty term at every voxel. The result captures 
pathologies because a DNN trained only on healthy data should exhibit high 
epistemic uncertainty in the region of a pathology; but areas of high 
aleatoric uncertainty may also have high epistemic uncertainty simply due to the 
network not being able to reliably estimate the corresponding regions (resulting 
from intrinsic properties of the data). The voxel-wise division retains areas of 
high epistemic uncertainty while reducing the intenstity of regions where the model 
just performs poorly. We can then use simple thresholding on the quotient to 
detect anomalies in test data. We call this novel quantity 
\emph{scibilic}\footnote{\emph{Scibilic} is an anglicized version of 
the Latin adjective \emph{scibilis}, which means knowable.} uncertainty because
it highlights the areas the model could know how to predict\textemdash provided 
sufficient representative training data\textemdash but does not.

\subsection{Network architecture}
\label{ssec:architecture}
We used a U-Net~\cite{ronneberger2015unet} architecture modified as follows:
\begin{itemize}[noitemsep]
\item We used two 3D convolutional layers, one at the start and one at the end. 
      This improved sharpness and slice-to-slice consistency.
\item We downsampled and upsampled three times instead of four. Experimental results
      showed no improvement with four downsample operations.
\item We substituted max-pooling layers for strided convolutions in
      downsampling. For upsampling we used nearest-neighbor interpolation followed by a $5^2$ convolution~\cite{odena2016deconvolution}. 
\item We attached two heads to the end of the network, where one output 
      $\hat{\mathbf{y}}$ and the other output $\hat{\boldsymbol{\sigma}}^2$. Both consisted
      of $3^3$ and $1^3$ convolutional layers.
\item We concatenated the input image to the feature maps output by the network immediately before both heads~\cite{zhao2017whole}.
\item We used spatial dropout~\cite{tompson2015efficient} ($p=0.2$)
      on all layers except the heads, 
      because it drops weights on convolutional layers
      unlike standard dropout~\cite{srivastava2014dropout}.
\item We used the AdamW optimizer \cite{loshchilov2019decoupled} with weight decay $10^{-6}$,
      learning rate 0.003, $\beta = (0.9,0.99)$, and batch size 36. 
\item We used $T=50$ weight samples in prediction.

\end{itemize}

\section{Results}
\label{sec:results}

\subsection{Dataset}

We used non-contrast $T_1$-w and CT images from 51 subjects on a protocol for 
retrospective data analysis approved by the institutional review board. Fifty of
the subjects were considered to be healthy, and the remaining subject
had anomalies in the brain and was excluded from training and validation. The
$T_1$-w images were acquired on a Siemens Magnetom Espree 1.5T
scanner (Siemens Medical Solutions, Erlangen, Germany, TE =
4.24~ms, TR = 1130~ms, flip angle = 15$^\circ$, image size =
512$\times$512~pixels, pixel size = 0.5$\times$0.5~mm$^2$, slice
thickness = 1~mm); geometric distortions were corrected on the Siemens
Syngo application. All $T_1$-w images were processed to normalize the
white matter mean \cite{reinhold2019evaluating}. The CT images were acquired on 
a Philips Brilliance Big Bore scanner (Philips Medical Systems, Andover, MA, image
size = 512$\times$512~pixels, pixel size = 0.6$\times$0.6~mm$^2$ –
0.8$\times$0.8~mm$^2$, slice thickness = 1.0~mm). All images
were resampled to have a digital resolution of $0.7 \times 0.7 \times
1.0$~mm$^3$. Finally, the $T_1$-w images were rigidly registered to the CT
images. For training, the $T_1$-w and CT images were split into
overlapping $128 \times 128 \times 8$~patches. We used 45 of the healthy subjects
for training and the remaining five healthy subjects for validation.
Test images were split into three overlapping segments along the 
inferior-superior axis due to memory constraints.

\begin{figure}[!t]
\centering
\begin{tabular}{ccccc}
\textbf{CT} & \textbf{Synthesized} & \textbf{Epistemic} & \textbf{Aleatoric} & \textbf{Scibilic} \\
\includegraphics[width=0.17\textwidth, clip, trim=0.2cm 0.2cm 0.2cm 1.2cm]{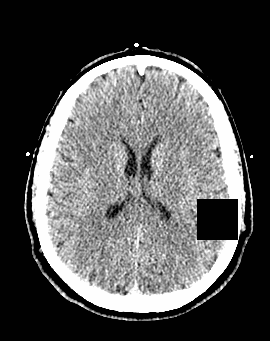} &
\includegraphics[width=0.17\textwidth, clip, trim=0.2cm 0.2cm 0.2cm 1.2cm]{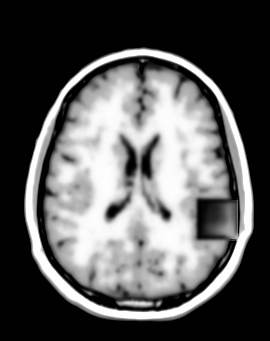} &
\includegraphics[width=0.17\textwidth, clip, trim=0.2cm 0.2cm 0.2cm 1.2cm]{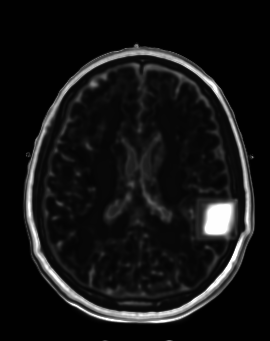} &
\includegraphics[width=0.17\textwidth, clip, trim=0.2cm 0.2cm 0.2cm 1.2cm]{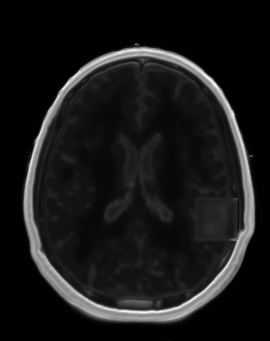} &
\includegraphics[width=0.17\textwidth, clip, trim=0.2cm 0.2cm 0.2cm 1.2cm]{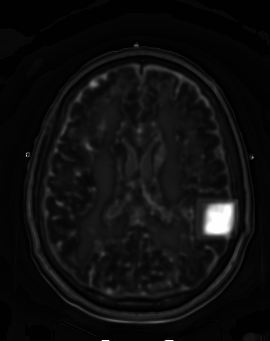}
\end{tabular}
\caption{\label{fig:epistemic}\textbf{Synthetic anomaly:}
Shown is an example synthetic anomaly CT image and the corresponding
estimated $T_1$-w image, epistemic, aleatoric, and scibilic (epistemic / aleatoric) uncertainty.}
\end{figure}

\begin{figure}[!b]
\centering
\includegraphics[width=0.45\textwidth, clip, trim=0cm 0mm 0cm 0mm]{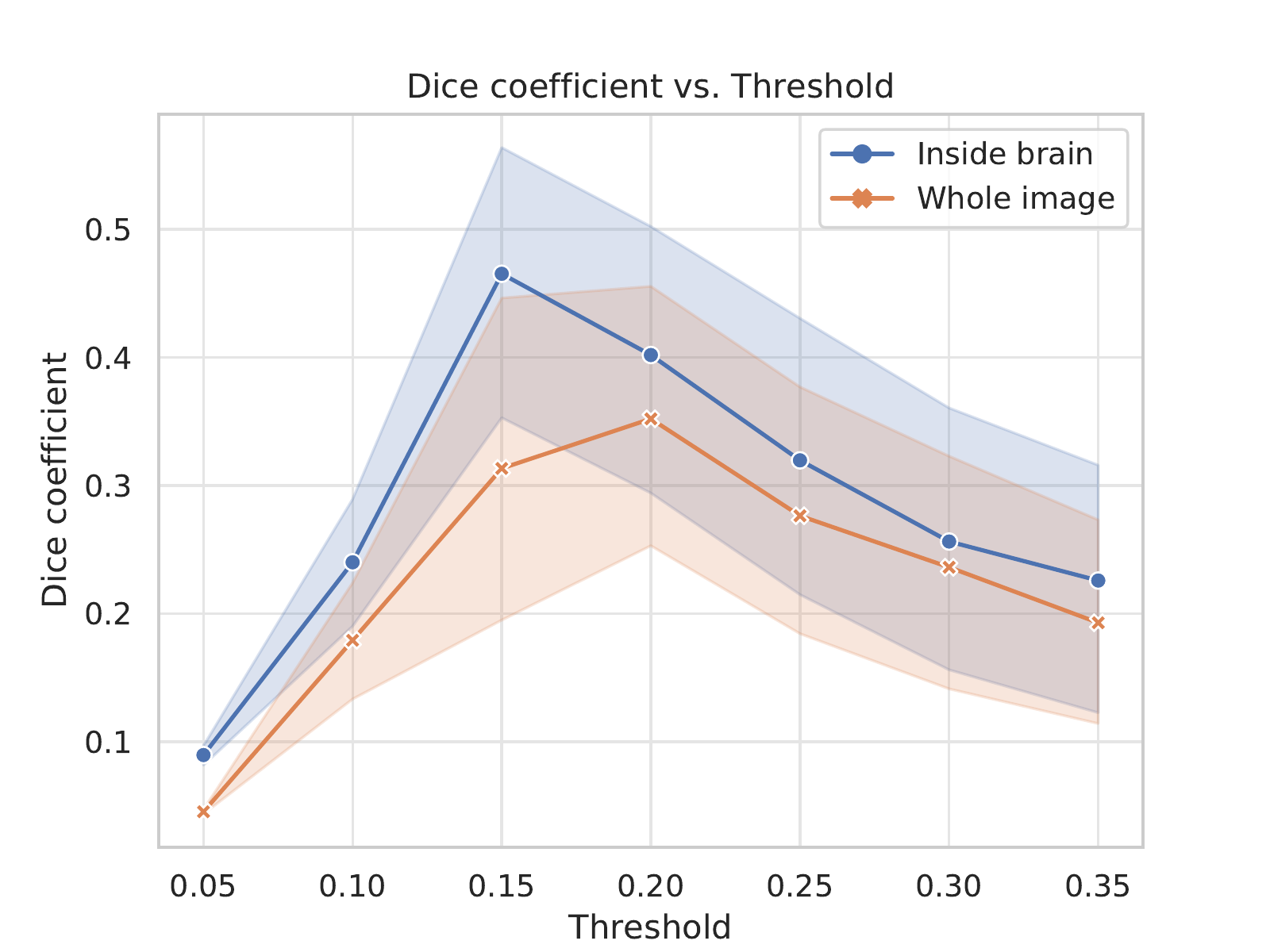}
\includegraphics[width=0.45\textwidth, clip, trim=0cm 0mm 0cm 0mm]{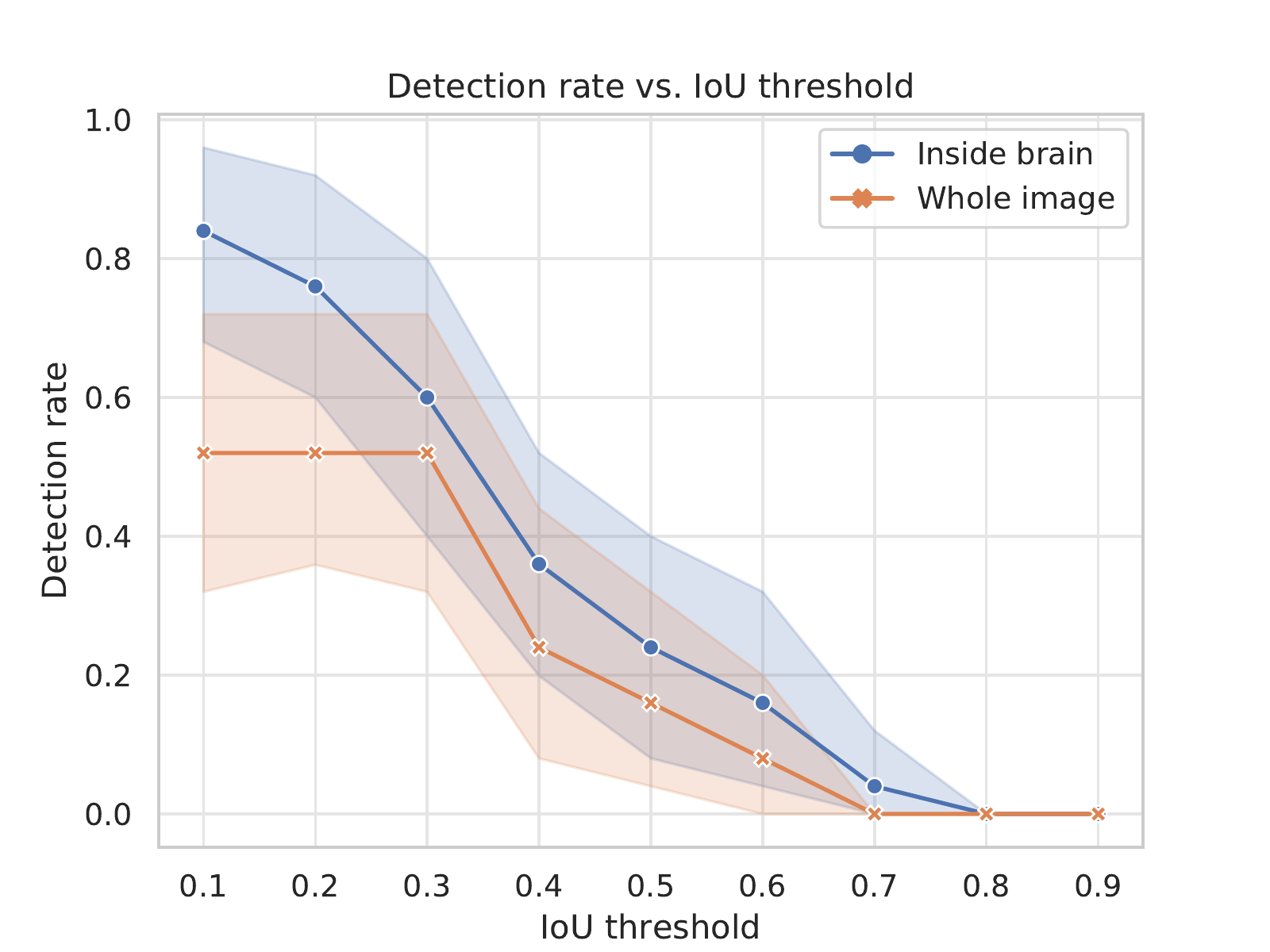}
\caption{\label{fig:quantep}\textbf{Dice coefficient on synthetic anomalies}: 
Shown are the Dice coefficients~(left) and detection rate~(right) calculated 
on the 25 test images. The threshold in the Dice coefficient plot corresponds to the
binarization threshold. The IoU threshold in the detection rate plot corresponds to
the IoU threshold used to determine detection. Shaded regions are 
bootstrapped 95\% confidence intervals.}
\end{figure}

\subsection{Unsupervised anomaly segmentation}

In our first experiment, we inserted synthetic anomalies into otherwise healthy
data\textemdash the five held-out validation datasets\textemdash and calculated the resulting Dice coefficient\cite{dice1945measures} and detection rate over a 
range of thresholds. Our synthetic anomaly is an all-zero cube, of 
side-length 40 voxels, placed randomly inside the brain mask of the five held-out 
healthy CT images (see Fig.~\ref{fig:epistemic} for an example). We created five of these 
anomalies per test subject by varying anomaly location. These 25 synthetic 
anomalous data were used as input to the trained network.
Quantitative results are in Fig.~\ref{fig:quantep}, where Intersection over Union 
(IoU) is used to indicate detection according to several IoU thresholds. Specifically, 
we calculated the IoU\textemdash also known as Jaccard index\cite{jaccard1912distribution}\textemdash for the 
largest-connected component of the binary prediction thresholded at 0.15. 

\begin{figure}[!t]
\centering
\begin{tabular}{ccccc}
\textbf{True} $T_1$ & \textbf{Synthesized} & \textbf{Epistemic} & \textbf{Aleatoric} & \textbf{Scibilic}\\
\includegraphics[width=0.17\textwidth, clip, trim=0.2cm 0.1cm 0.2cm 1.4cm]{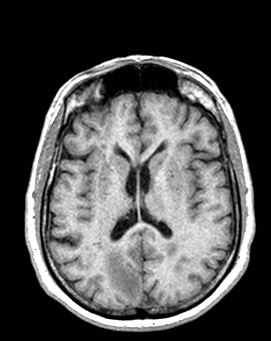} &
\includegraphics[width=0.17\textwidth, clip, trim=0.2cm 0.1cm 0.2cm 1.4cm]{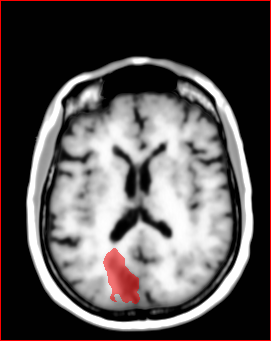} &
\includegraphics[width=0.17\textwidth, clip, trim=0.2cm 0.1cm 0.2cm 1.4cm]{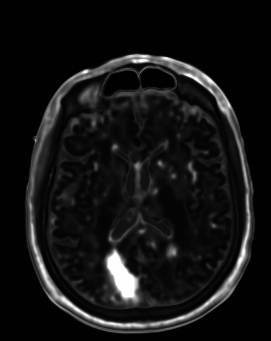} &
\includegraphics[width=0.17\textwidth, clip, trim=0.2cm 0.1cm 0.2cm 1.4cm]{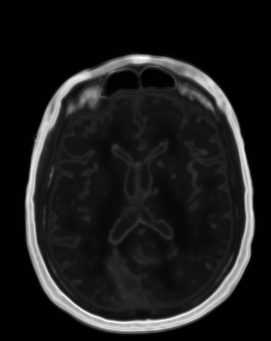} &
\includegraphics[width=0.17\textwidth, clip, trim=0.2cm 0.1cm 0.2cm 1.4cm]{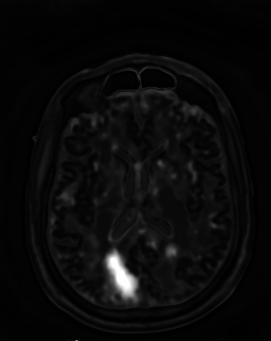}
\end{tabular}
\caption{\label{fig:comp_ex}\textbf{Example anomalous image:}
Shown is an example anomalous image and corresponding synthesized image along 
with the estimated epistemic and aleatoric uncertainty maps as well as their 
voxel-wise quotient (scibilic). The red overlay in the synthesized image is the largest-connected
component of the binarized scibilic uncertainty image at a threshold of 0.15, which was the optimal
threshold in the synthetic anomaly experiment.}
\end{figure}

In our second experiment, we tested the trained network on the held-out 
pathological dataset (pathology in the occipital lobe) collected on the 
same scanner. Since we do not have a ground-truth label for this data set we
show qualitative results in Fig.~\ref{fig:comp_ex}. 

\section{Discussion and Conclusion}
\label{sec:conclusion}

Detecting and segmenting pathologies in structural medical images is difficult, 
especially without labels. In the experiments above, we showed that measures of 
uncertainty can be used to highlight anomalous regions in images simply by
training the proposed network on healthy data alone. Importantly, this 
anomalous region segmentation is done in conjunction with a CT-to-MR image translation task
that provides more insight into the fine anatomical structures of the subject.
While the resulting synthesized image is not state-of-the-art, the additional 
information provided by the estimates of uncertainty provides researchers with a 
greater ability to understand the model and data as well as detect and segment 
anomalies in test images.

\acknowledgments
This work was supported by 12 Sigma Technologies.

\bibliographystyle{spiebib}
\IfFileExists{/Users/jcreinhold/Research/code/mr-synth/notes/roadmap/biblio.bib}%
{\bibliography{/Users/jcreinhold/Research/code/mr-synth/notes/roadmap/biblio.bib}}
{\bibliography{biblio}}

\end{document}